\documentclass[%
reprint,
superscriptaddress,
showpacs,
amsmath,amssymb,
aps,
prc,
showkeys,
twocolumn,
]{revtex4-1}

\usepackage{graphicx}% Include figure files
\usepackage{dcolumn}% Align table columns on decimal point
\usepackage{bm}% bold math
\usepackage{hyperref}% add hypertext capabilities
\hypersetup{colorlinks=true, citecolor=blue, urlcolor=blue, linkcolor=blue}
\usepackage{color}
\usepackage{epsfig}
\usepackage{epstopdf}
\def\nuc#1#2{\relax\ifmmode{}^{#1}{\protect\text{#2}}\else${}^{#1}$#2\fi}

\begin{document}
  \newcommand {\nc} {\newcommand}
  \nc {\beq} {\begin{eqnarray}}
  \nc {\eeq} {\nonumber \end{eqnarray}}
  \nc {\eeqn}[1] {\label {#1} \end{eqnarray}}
  \nc {\eol} {\nonumber \\}
  \nc {\eoln}[1] {\label {#1} \\}
  \nc {\ve} [1] {\mbox{\boldmath $#1$}}
  \nc {\ves} [1] {\mbox{\boldmath ${\scriptstyle #1}$}}
  \nc {\mrm} [1] {\mathrm{#1}}
  \nc {\half} {\mbox{$\frac{1}{2}$}}
  \nc {\thal} {\mbox{$\frac{3}{2}$}}
  \nc {\fial} {\mbox{$\frac{5}{2}$}}
  \nc {\la} {\mbox{$\langle$}}
  \nc {\ra} {\mbox{$\rangle$}}
  \nc {\etal} {\emph{et al.}}
  \nc {\eq} [1] {(\ref{#1})}
  \nc {\Eq} [1] {Eq.~(\ref{#1})}
  \nc {\Ref} [1] {Ref.~\cite{#1}}
  \nc {\Refc} [2] {Refs.~\cite[#1]{#2}}
  \nc {\Sec} [1] {Sec.~\ref{#1}}
  \nc {\chap} [1] {Chapter~\ref{#1}}
  \nc {\anx} [1] {Appendix~\ref{#1}}
  \nc {\tbl} [1] {Table~\ref{#1}}
  \nc {\Fig} [1] {Fig.~\ref{#1}}
  \nc {\ex} [1] {$^{#1}$}
  \nc {\Sch} {Schr\"odinger }
  \nc {\flim} [2] {\mathop{\longrightarrow}\limits_{{#1}\rightarrow{#2}}}
  \nc {\IR} [1]{\textcolor{red}{#1}}
  \nc {\IB} [1]{\textcolor{blue}{#1}}
  \nc{\pderiv}[2]{\cfrac{\partial #1}{\partial #2}}
  \nc{\deriv}[2]{\cfrac{d#1}{d#2}}

\title{$^{15}$C: from halo effective field theory structure\\ to the study of transfer, breakup and radiative-capture reactions}%
\author{Laura Moschini}
\email{laura.moschini@ulb.ac.be}
\affiliation{Physique Nucl\'eaire et Physique Quantique (C.P.~229)\\
Universit\'e libre de Bruxelles (ULB), 50 avenue F.D.\ Roosevelt, B-1050 Brussels, Belgium}
\author{Jiecheng Yang}
\email{jiecyang@ulb.ac.be}
\affiliation{Physique Nucl\'eaire et Physique Quantique (C.P.~229)\\
Universit\'e libre de Bruxelles (ULB), 50 avenue F.D.\ Roosevelt, B-1050 Brussels, Belgium}
\affiliation{Afdeling Kern-en Stralingsfysica, Celestijnenlaan 200d-bus 2418, 3001 Leuven, Belgium}
\author{Pierre Capel}
\email{pcapel@uni-mainz.de}
\affiliation{Institut f\"ur Kernphysik, 
Johannes Gutenberg-Universit\"at Mainz,
Johann-Joachim-Becher Weg 45, 
D-55099 Mainz, Germany}
\affiliation{Physique Nucl\'eaire et Physique Quantique (C.P.~229)\\
Universit\'e libre de Bruxelles (ULB), 50 avenue F.D.\ Roosevelt, B-1050 Brussels, Belgium}

\date{\today}

\begin{abstract}
\begin{description}
\item[Background] 
Aside from being a one-neutron halo nucleus, $^{15}$C is interesting because it is involved in reactions of relevance for several nucleosynthesis scenarios.
\item[Purpose]
The aim of this work is to analyze various reactions involving $^{15}$C, using a single structure model based on halo effective field theory (Halo EFT) following the excellent results obtained in [P.~Capel, D.~R.~Phillips, and H.-W.~Hammer, Phys. Rev. C {\bf 98}, 034610 (2018)].
\item[Method] 
To develop a Halo-EFT model of $^{15}$C at next to leading order (NLO), we first extract the asymptotic normalization coefficient (ANC) of its ground state by analyzing $^{14}{\rm C}(d,p)^{15}{\rm C}$ transfer data at low energy using the method developed in [J. Yang and P. Capel, Phys. Rev. C {\bf 98}, 054602 (2018)].
Using the Halo-EFT description of $^{15}$C constrained with this ANC, we study the $^{15}$C Coulomb breakup at high (605~MeV/nucleon) and intermediate (68~MeV/nucleon) energies using eikonal-based models with a consistent treatment of nuclear and Coulomb interactions at all orders, and which take into account proper relativistic corrections. 
Finally, we study the $^{14}{\rm C}(n,\gamma)^{15}{\rm C}$ radiative capture.
\item[Results] 
Our theoretical cross sections are in good agreement with experimental data for all reactions, thereby assessing the robustness of the Halo-EFT model of this nucleus. 
Since a simple NLO description is enough to reproduce all data, the only nuclear-structure observables that matter are the $^{15}$C binding energy and its ANC, showing that all the reactions considered are purely peripheral.
In particular, it confirms the value we have obtained for the ANC of the $^{15}$C ground state: ${\cal C}^2_{1/2^+}=1.59\pm0.06$~fm$^{-1}$. 
Our model of $^{15}$C provides also a new estimate of the radiative-capture cross section at astrophysical energy: $\sigma_{n,\gamma}(23.3~{\rm keV})=4.66\pm0.14~\mu$b.

\item[Conclusions]
Including a Halo-EFT description of $^{15}$C within precise models of reactions is confirmed to be an excellent way to relate the reaction cross sections and the structure of the nucleus.
Its systematic expansion enables us to establish how the reaction process is affected by that structure and deduce which nuclear-structure observables are actually probed in the collision.
From this, we can infer valuable information on both the structure of $^{15}$C and its synthesis through the $^{14}{\rm C}(n,\gamma)^{15}{\rm C}$ radiative capture at astrophysical energies.
\end{description}
\end{abstract}
%\pacs{23.23.+x, 56.65.Dy}
%%\keywords{breakup; relativistic correction} %%showkeys
\maketitle

\section{Introduction}
The nucleus $^{15}$C is interesting for various reasons.
On a nuclear-structure viewpoint, $^{15}$C is one of the best known one-neutron halo nuclei \cite{Tan96,Rii13}.
Due to its small one-neutron separation energy [$S_n(^{15}{\rm C})=1.218$~MeV], 
the ground state of $^{15}$C is mostly described as a two-body structure, 
in which the valence neutron is loosely bound in a $1s_{1/2}$ orbital to a $^{14}$C in its $0^+$ ground state.
Thanks to its loose binding and the fact that it sits in an $l=0$ orbital, the valence neutron exhibits a high probability of presence at a large distance from the other nucleons.
It therefore forms like a diffuse halo surrounding a compact core \cite{HJ87}.
The existence of halos in some nuclei challenges our view of the nucleus, which is usually seen as a compact object with a nucleon density at saturation.
Halo nuclei, including $^{15}$C, are thus the focus of many experimental and theoretical studies \cite{Tan96,Rii13}.

The study of $^{15}$C has also applications in nuclear astrophysics.
Its synthesis through one-neutron radiative capture by $^{14}$C has been suggested to be part of neutron-induced CNO cycles, which take place in the helium-burning zone of asymptotic-giant-branch (AGB) stars \cite{WGS99}.
This $^{14}$C$(n,\gamma)^{15}$C reaction is also the doorstep to the production of heavy elements in inhomogeneous big-bang nucleosynthesis \cite{KMF90} and it has been shown to be part of possible reaction routes in the nuclear chart during the $r$ process in Type II supernov\ae\ \cite{Ter01}.
It is therefore necessary to have a reliable estimate of the cross section for this radiative capture at astrophysical energy, and hence to better understand the structure of $^{15}$C.

Because $^{15}$C exhibits a short lifetime, its structure cannot be probed with usual spectroscopic techniques.
This nucleus is therefore mostly studied through reactions.
Transfer, such as $(d,p)$, measured in both direct and inverse kinematics, has been used to infer the single-particle structure of $^{15}$C \cite{14MeVdp,Cec75,MSD94,17MeVdp}.
In breakup, the lose binding of the valence neutron to the core is broken up during the collision of the nucleus on a target, hence revealing its internal core-$n$ structure.
Various experimental campaigns have been set up to measure the inclusive breakup---also known as knockout---of $^{15}$C on light targets at intermediate beam energies \cite{Tos02,Sau04,Fang04}.
In these measurements, only $^{14}$C is detected after the reaction, and information pertaining to the single-particle structure of $^{15}$C is inferred from the analysis of the parallel-momentum distribution of the core.
In Refs.~\cite{Dat03,Nak09-15C}, the Coulomb (exclusive) breakup of $^{15}$C has been measured.
In that case, both the $^{14}$C core and the halo neutron are detected in coincidence after the dissociation of the $^{15}$C projectile on a Pb target.
Being dominated by the Coulomb interaction, this reaction process is rather clean as it exhibits little dependence on the choice of the optical potentials used to describe the nuclear interaction between the projectile constituents (core and $n$) with the target.

In addition to its interest in the study of the halo structure of $^{15}$C, Coulomb breakup has also been suggested as an indirect method to deduce the cross section for the $^{14}{\rm C}(n,\gamma)^{15}$C radiative capture at low energies \cite{BBR86,BHT03}.
The idea behind the Coulomb-breakup method is that this dissociation, which is often described as resulting from the exchange of virtual photons between the projectile and the heavy target \cite{winther-alder}, can be seen as the time-reversed reaction of the radiative capture, where a (real) photon is emitted following the capture of a neutron by the core.
Later analyses have shown that the breakup process is not that simple and that higher-order effects spoil this nice picture \cite{EBS05,CB05}.
However, it has been suggested that the Coulomb-breakup measurements could be used to infer the asymptotic normalization coefficient (ANC) of the $^{15}$C ground-state wave function \cite{SN08,SN08err}.
However, due to the aforementioned higher-order effects, a precise model of the reaction is needed in the analysis of the reaction \cite{SN08,Esb09,CN17}.
Because the radiative capture $^{14}{\rm C}(n,\gamma)^{15}$C is a purely peripheral process \cite{15CANC_2006}, a reliable estimate of this ANC can then be used to compute its cross section.
Following \Ref{15CANC_2006}, it has also been suggested to rely on the strong sensitivity of transfer reaction to the single-particle structure of the nucleus to measure the ANC of the $^{15}$C ground-state wave function for that purpose \cite{17MeVdp}.

Since the radiative capture $^{14}{\rm C}(n,\gamma)^{15}$C has been measured directly by Reifarth \etal\ \cite{Reifarth2008}, the $^{15}$C case provides the opportunity to test the validity of the different indirect methods listed above.

In the present work, we reanalyze the transfer \cite{14MeVdp,17MeVdp}, Coulomb-breakup \cite{Dat03,Nak09-15C} and radiative-capture \cite{Reifarth2008} measurements using one single description of the one-neutron halo nucleus $^{15}$C.
For this, we follow the recent idea developed in \Ref{CPH18} 
and include, within precise models of reactions, 
a description of the nucleus based on halo effective field theory (Halo EFT) \cite{BHK02} 
(see Ref. \Ref{HJP17} for a recent review).
Halo EFT exploits the natural separation of scales that is observed in halo nuclei---viz. the difference between the small size of the core $R_{\rm core}$ and the large extension of the halo $R_{\rm halo}$---to build an effective Hamiltonian constructed as an expansion in powers of the small parameter $R_{\rm core}/R_{\rm halo}$.
This allows us to introduce, order by order, the different nuclear-structure parameters in the description of the nucleus within the reaction models, and thereby to deduce how each of them affects the reaction processes.
This puts a strong constraint on what can be learned about the structure of $^{15}$C from transfer and breakup experiments and how this nuclear-structure information relates to the direct radiative-capture capture measurement of \Ref{Reifarth2008}.

This article is structured as follows. 
In \Sec{sec2} we introduce the Halo EFT description of $^{15}$C and explain how it is fitted at next to leading order (NLO).
Using this description, we reanalyze transfer measurements at $E_d=14$  \cite{14MeVdp} and $17.06$~MeV  \cite{17MeVdp} in Sec.~\ref{transf-sec}.
In \Sec{sec3} we use the same $^{15}$C structure to study its breakup at high (605~MeV/nucleon \cite{Dat03}) and intermediate (68~MeV/nucleon \cite{Nak09-15C}) energy. 
In \Sec{sec4}, we study the $^{14}{\rm C}(n,\gamma)^{15}{\rm C}$ radiative capture \cite{Reifarth2008}.
Finally, in \Sec{sec5}, we summarize our results and provide the outlook for future work.

\section{Halo-EFT description of $^{15}$C \label{sec2}}
\subsection{Single-particle structure of $^{15}$C \label{15Cstructure}}
Being a one-neutron halo nucleus, $^{15}$C can be modeled as a neutron loosely bound to a $^{14}$C core.
With the assumption that the $^{14}$C core is in its ground state ($0^+$),  the $\half^+$ ground state (g.s.) of  $^{15}$C can be described by a $^{14}$C($0^+$)$\otimes 1s_{1/2}$ configuration  and its $\fial^+$ excited state (e.s.) by a $^{14}$C($0^+$)$\otimes 0d_{5/2}$.
These states have an energy relative to the one-neutron threshold of $E_{\rm g.s.} = -1.218$~MeV and $E_{\rm e.s.} = -0.478$~MeV, respectively.

To model this system, the core $A$ of mass $m_A$ and charge $Z_A e$ is assumed to be of spin and parity $0^+$ and we neglect its internal structure.
The halo nucleus $B=A+n$ is thus of mass $m_B=m_A+m_n$, with $m_n$ the neutron mass, and charge $Z_B e=Z_A e$.
Such a two-body structure is described by the internal Hamiltonian
\beq
H_0=-\frac{\hbar^2\Delta}{2\mu_{An}} + V_{An}(\ve{r}),
\eeqn{e0}
where $\ve{r}$ is the $A$-$n$ relative coordinate, $\mu_{An} = m_A m_n /m_B$ is their reduced mass, and $V_{An}$ is the effective potential simulating their interaction.
In partial wave $ljm$, the eigenstates of $H_0$ read
\beq
H_0\ \varphi_{ljm}(E_{lj},\ve{r}) = E_{lj}\ \varphi_{ljm}(E_{lj},\ve{r}),
\eeqn{e0a}
where $j$ is the total angular momentum resulting from the coupling of the orbital angular momentum $l$ with the spin of the halo neutron and $m$ is its projection.
The eigenstates of $H_0$ of negative energy $E_{n'lj}$ are discrete and correspond to the bound states of the 
two-body model of the projectile $B$. These include physical $A$-$n$ bound states of the system 
as well as Pauli forbidden states, which simulate the presence of neutrons within the core $A$.
We enumerate them by adding the number of nodes in the radial wave function $n'$ to the other quantum numbers.
They are normed to unity and their reduced radial wave function behaves asymptotically as
\beq
u_{n'lj}(r)\flim{r}{\infty}b_{n'lj}\, ik_{n'lj}r\ h_l^{(1)}(ik_{n'lj} r),
\eeqn{e0b}
where $\hbar k_{n'lj}=\sqrt{2\mu_{An}|E_{n'lj}|}$, whith $|E_{n'lj}|$ the $A$-$n$ binding energy, and $h_l^{(1)}$ is a spherical Bessel function of the third kind \cite{AS70}.
The single-particle asymptotic normalization constant (SPANC) $b_{n'lj}$ defines the
strength of the exponential tail of the $A$-$n$ bound-state wave function \cite{ANC77}. 
This SPANC will vary with the geometry of the potential used to simulate the $A$-$n$ interaction \cite{15CANC_2014,BPODEG14,Tim14,CN06}.
The asymptotic behavior (\ref{e0b}) is universal, therefore it exists also in the actual structure of the nucleus, viz.\ in the overlap wave function obtained within a microscopic calculation of the nucleus \cite{CDN10,Tim14}. 
Being affected by the inherent couplings between the different configurations in the actual structure of the nucleus, in particular those involving the core in one of its excited states, the true asymptotic normalization constant (ANC) of the overlap wave function of the physical state of spin and parity $J^\pi$ corresponding to the configuration in which the core is in its $0^+$ ground state, ${\cal C}_{J^\pi}$, differs from the SPANC $b_{n'lj}$ obtained in the effective single-particle description considered here \cite{CDN10,Tim14}.

The positive-energy states describe the $A$-$n$ continuum, i.e.\ the broken-up projectile.
Their reduced radial parts are normalized according to
\beq
u_{klj}\flim{r}{\infty} kr\left[\cos \delta_{lj}\ j_l(kr) + \sin\delta_{lj}\ n_l(kr)\right]
\eeqn{e0c}
where $\delta_{lj}$ is the phaseshift at energy $E_{lj}$ and $\hbar k = \sqrt{2\mu_{An}E_{lj}}$; 
$j_l$ and $n_l$ are spherical Bessel functions of the first and second kinds, respectively \cite{AS70}.

As mentioned above, the $A$-$n$ interaction is described by an effective potential $V_{An}$.
In this study, following the idea developed in \Ref{CPH18}, this potential is built within a Halo-EFT 
description of the nucleus \cite{BHK02,HJP17}.
At the leading order (LO), this interaction consists of a simple contact term within the sole $s$ wave.
As usual, this interaction is regularized with a Gaussian
\beq
V_{An}^{\rm LO}(r) = V_0^{s1/2} e^{-\frac{r ^2}{2 r_0 ^2}}.
\eeqn{e-LO}
The range of the Gaussian $r_0$ corresponds to the scale of the short-range physics neglected in this Halo-EFT description.
Changing its value will enable us to generate different single-particle wave functions to describe the 
$^{14}$C-$n$ system and hence test the sensitivity of our reaction calculations to the internal part of the wave function of the projectile.
At LO, the only free parameter $V_0^{s1/2}$ is adjusted to reproduce $E_{\rm g.s.}=-1.218$~MeV within a $1s_{1/2}$ orbit.

At next-to-leading order (NLO), the interaction is extended up to the $p$ waves and contains, 
in addition to the contact term its second-order derivative.
For simplicity, we follow Ref.~\cite{CPH18}
and use the equivalent following parametrisation of the interaction
\beq
V_{An}^{\rm NLO}(r) = V_0^{lj} e^{-\frac{r ^2}{2 r_0 ^2}}+V_2^{lj} r^2e^{-\frac{r ^2}{2 r_0 ^2}}.
\eeqn{e-1_1}
To constrain the potential parameters $V_0^{s1/2}$ and $V_2^{s1/2}$ in the $s$ wave, we need two structure observables: in addition to the binding energy of the state, we also use its ANC. 
Various groups have estimated this ANC from reaction data \cite{15CANC_2002,15CANC_2006,15CANC_2007,SN08,SN08err,17MeVdp,15CANC_2014}. 
In this work, we use the method presented in \Ref{yang-capel} to deduce this ANC from low-energy transfer data selected at forward angle (see \Sec{transfer}). 

Unlike $^{11}$Be, $^{15}$C does not exhibit any low-lying bound or resonant $\thal^-$ or $\half^-$  states to which we could fit the effective interaction \eq{e-1_1} in the $p$ waves.
Therefore, true to the spirit of Halo-EFT, we set this interaction to 0 in the $p_{3/2}$ and $p_{1/2}$ partial waves.
Interestingly, this treatment is in agreement with preliminary results obtained in an \emph{ab initio} calculation of $^{15}$C performed within the no-core shell model with continuum (NCSMC), which predicts negligible phaseshifts at low $^{14}$C-$n$ energies in both $p$ waves \cite{Nav18p}.
%The parameter $r_0$ is related to the short-range physics of the nucleus.
%Since there are no direct measurement of ANC, many groups have been analyzing different experiments involving the nucleus under examination, and have given their theoretical prediction \cite{15CANC_2002,15CANC_2006,15CANC_2007,15CANC_2008,17MeVdp,15CANC_2014}. 
%These values are listed in Tab.\ \ref{tab3} in Sec.\ \ref{transfer}.
%In this work, we use the method presented in Ref.\ \cite{yang-capel} to get the ANC of from transfer measurements. 

At NLO, the interaction $V_{An}$ is nil in higher partial waves.
Since the $\fial^+$ excited bound state of $^{15}$C plays a role in the radiative capture (see \Sec{sec4}), we follow the idea of \Ref{CPH18} and go beyond NLO to include a $0d_{5/2}$ state at $E_{\rm e.s.}=-0.478$~MeV.
The potential in that partial wave is chosen similar to that of \Eq{e-1_1}.
We fit the depths $V_{0}^{d5/2}$ and $V_{2}^{d5/2}$ to reproduce the experimental binding energy of the $\fial^+$ state and the ANC deduced from transfer data.

\subsection{Extraction of the ANC of the $^{15}$C bound states from the analysis of low-energy transfer reactions \label{transfer}}
To obtain a reliable estimate of the ANC of both bound states of $^{15}$C, we follow the idea developed in Ref.~\cite{yang-capel} and reanalyze $^{14}$C$(d,p)^{15}$C transfer data.
In that reference, it was found that $(d,p)$ transfer reactions are purely peripheral when they are performed at low beam energy (viz. $E_d\lesssim15$~MeV) and when the data are selected at forward angles.
Within these experimental conditions, the transfer cross section scales perfectly with the square of the final state ANC ${\cal C}_{J^\pi}^2$.
That value can then be reliably extracted from a comparison between reaction calculations performed using a single-particle description of the nucleus similar to the one presented in \Sec{15Cstructure} and experimental data \cite{yang-capel}.

We therefore need $^{14}$C$(d,p)^{15}$C transfer data measured at low energies, and which contain enough data points at forward angles for this extraction of the ANC of $^{15}$C to be statistically meaningful.
Two experiments satisfying the low-energy condition have been performed: one at the University of Notre Dame at $E_d=14$~MeV \cite{14MeVdp}, and another at the Nuclear Physics Institute of the Czech Academy of Sciences at $E_d=17.06$~MeV \cite{17MeVdp}.
Unfortunately, the former contains only one point at $\theta<15^\circ$, which we deem not enough for this extraction.
Fortunately, although performed at a slightly higher energy, the latter experiment contains six points at $\theta<12^\circ$, which seems enough to constrain the ANC within proper peripheral conditions (see below).

Following the method presented in \Ref{yang-capel}, we couple a leading order (LO) Halo-EFT description of $^{15}$C with a finite-range adiabatic distorted wave approximation (FR-ADWA) model \cite{JT74}.
This model provides a reliable description of transfer reactions at these energies \cite{NunesDeltuva2011,Upadhyay2012}.
As in \Ref{yang-capel}, we consider the CH89 global potential \cite{VTMLC91} to generate the optical potentials in the incoming ($d$-$^{14}$C) and outgoing ($p$-$^{15}$C) channels.
The Reid soft core potential \cite{Reid68} is used to compute the deuteron bound state. 
The deuteron adiabatic potentials are obtained with the frontend code of TWOFNR \cite{IT08}
and the transfer calculations are performed using FRESCO \cite{FRESCO}.
We illustrate here the results for the ground state, the method to extract the ANC of the excited state is analogous, though less efficient because it corresponds to a $d$ $^{14}$C-$n$ bound state (see Ref.~\cite{yang-capel} for the details).

\begin{figure}
\includegraphics[width=\columnwidth]{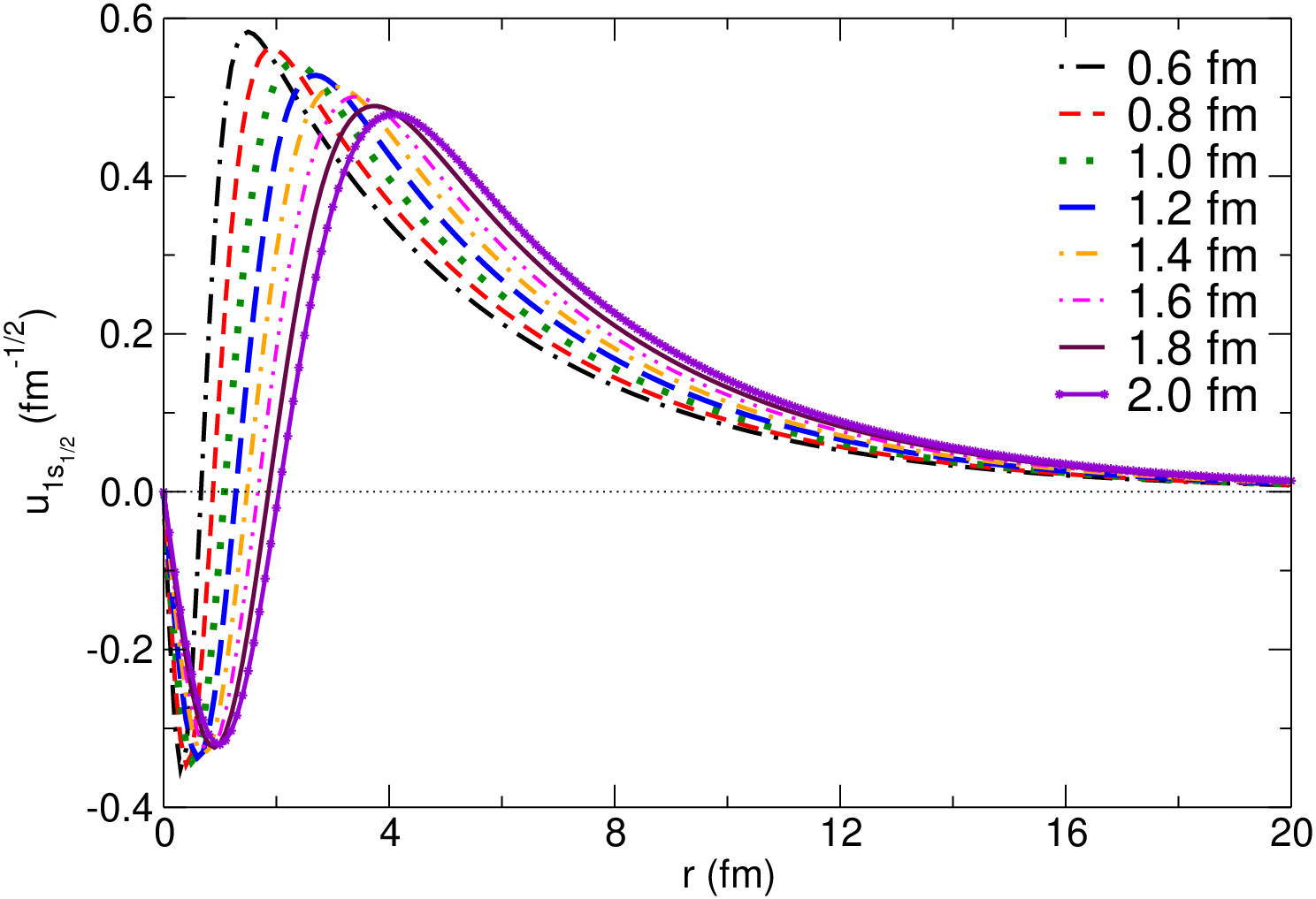}
\caption{\label{f-0_17MeV} Reduced radial wave functions of the  $^{15}$C g.s.\ 
obtained with LO Gaussian potentials of Table \ref{tab2}.}
\end{figure}

We first build eight Gaussian potentials at the LO of Halo-EFT [see \Eq{e-LO}] considering different ranges $r_0$ between $0.6$~fm and $2.0$~fm. 
For each width the depth $V_0^{s1/2}$ is adjusted to reproduce the neutron binding energy in the $^{15}$C final state (see Table \ref{tab2}).
These potentials provide different single-particle radial wave functions $u_{1s1/2}$ 
with very different SPANCs $b^{(r_0)}_{1s1/2}$, but also a significant change in the surface part of the nucleus, 
i.e.\ in the range $2~{\rm fm}\lesssim r\lesssim 4$~fm, see \Fig{f-0_17MeV}.
This is the corner stone of the method developed in \Ref{yang-capel}, because it is known that transfer reactions can be sensitive to that region \cite{15CANC_2007,Tim14}.
Using single-particle wave functions that strongly differ, not only in their SPANC, but also in their shape within that surface region will enable us to accurately determine the conditions under which the reaction is purely peripheral, and thus under which a reliable estimate of the actual ANC of the nucleus can be inferred.
\begin{table}[h!]
\begin{tabular}{ccc}
$r_0$ (fm) & ~$V_0^{s1/2}$ (MeV)~ & $b^{(r_0)}_{1s1/2}$ (fm$^{-1/2}$) \\
\hline
0.6 & -591.05 & 0.865\\
0.8 & -339.87 & 0.934\\
1.0 & -222.43 & 1.01\\
1.2 & -157.95 & 1.09\\
1.4 & -118.68 & 1.17\\
1.6 & -92.933 & 1.26\\
1.8 & -75.095 & 1.36\\
2.0 & -62.212  & 1.46\\
\hline
\end{tabular}
\caption{Potentials describing $^{14}$C$+n$ g.s.\ at LO [see Eq.~(\ref{e-LO})] 
and corresponding single-particle asymptotic normalization constant (SPANC) $b^{(r_0)}_{1s1/2}$.
They are adjusted on the one-neutron binding energy.}
\label{tab2}
\end{table}

\begin{figure}
\includegraphics[width=\columnwidth]{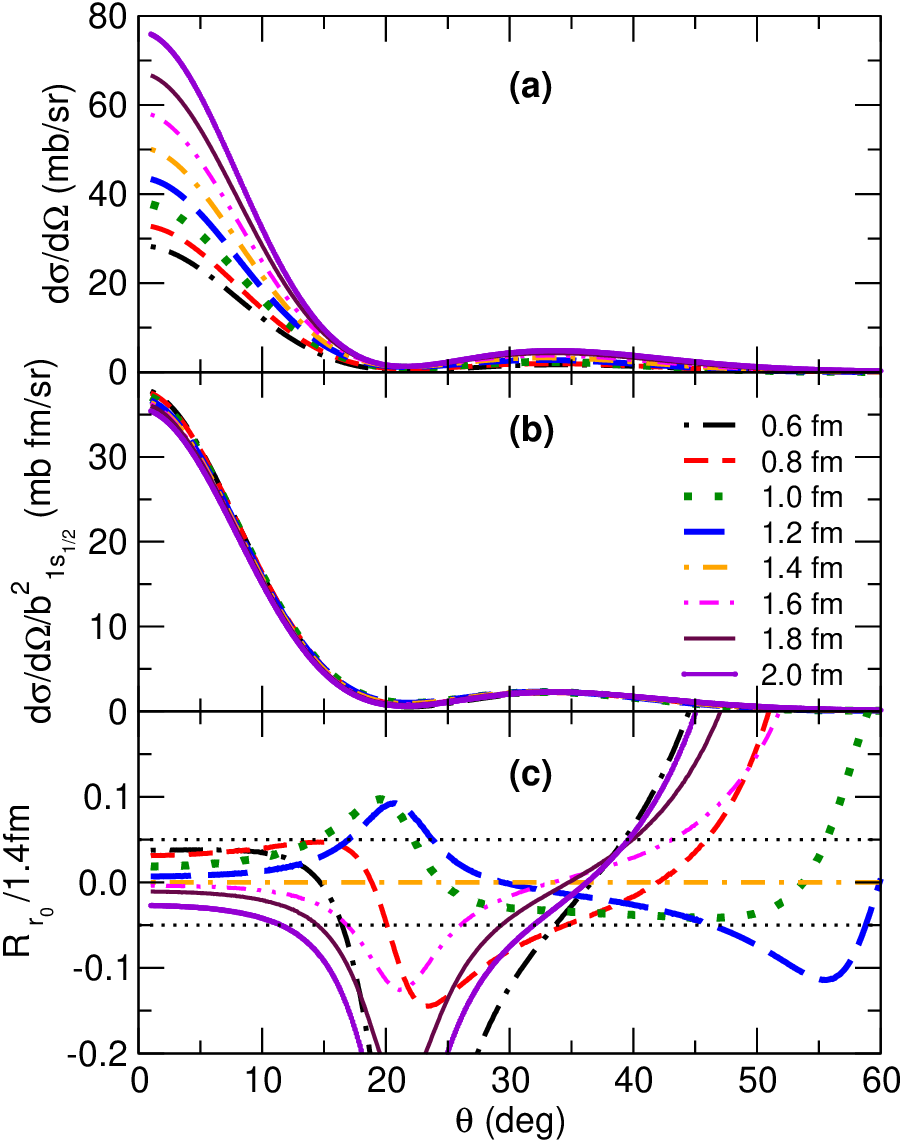}
\caption{\label{f-1_17MeV} Analysis of the differential cross section of $^{14}$C$(d,p)^{15}$C(g.s.) for the deuteron energy 
$E_d=17.06$~MeV. The results of the FR-ADWA calculations are presented for every wave function of Fig.\ \ref{f-0_17MeV}.}
\end{figure}

With this input, we compute within the FR-ADWA \cite{JT74} the corresponding theoretical differential cross section $d\sigma_{\rm th}/d\Omega$ for the transfer to the $^{15}$C g.s.\ at $E_d=17.06$~MeV \cite{17MeVdp}, expressed as a function of the relative direction $\Omega = (\theta, \phi)$ between the proton and the $^{15}$C in the outgoing channel.
These results are displayed in \Fig{f-1_17MeV}(a) for the eight g.s.\ wave functions shown in \Fig{f-0_17MeV}.
At forward angles, the cross sections exhibit a huge sensitivity to the choice of the $^{14}$C-$n$ wave function.
They seem to scale with the square of the SPANC, as one would expect if the process were purely peripheral \cite{yang-capel}. 
To confirm this,  we have plotted the transfer cross section scaled by $b_{1s1/2}^2$ in \Fig{f-1_17MeV}(b). 
In this way, the spread in the results is significantly reduced at forward angles. 

To precisely determine within which angular range the data should be limited to select strictly peripheral conditions, we remove the major angular dependence by considering the ratio
\beq
{\cal R}_{r_0/1.4\,\rm fm}(\theta) = \left( \frac{b^{(1.4\,\rm fm)}_{n'lj}}{b^{(r_0)}_{n'lj}}  \right)^2 \frac{d\sigma^{(r_0)}_{\rm th}/d\Omega}{d\sigma^{(1.4\,\rm fm)}_{\rm th}/d\Omega}-1,
\eeqn{e-1_2}
where the transfer cross section computed using the $^{14}$C-$n$ Gaussian potential of range $r_0$, scaled by the square of the corresponding SPANC $b_{1s1/2}^{(r_0)}$, is divided by the result obtained with $r_0 = 1.4$~fm, which is at the center of the range in $r_0$.
The results are displayed in \Fig{f-1_17MeV}(c).
We see that all ratios ${\cal R}_{r_0/1.4\,\rm fm}$ fall very close to one another at small angles, confirming the peripherality of the reaction when data measured at low beam energy are selected in the forward direction.
To define an angular range in which the reaction can be considered
as peripheral, we consider a maximum of $5\%$ difference [horizontal black dotted lines in \Fig{f-1_17MeV}(c)]. 
In this case, this happens only at very forward angles, viz.\ when $\theta < 12^{\circ}$.
There are six data points within this angular region in this experiment \cite{17MeVdp}.
Note that there is no data available within this angular range in the case of the experiment performed at the lower energy $E_d=14$~MeV \cite{14MeVdp}.

\begin{figure}
\includegraphics[width=\columnwidth]{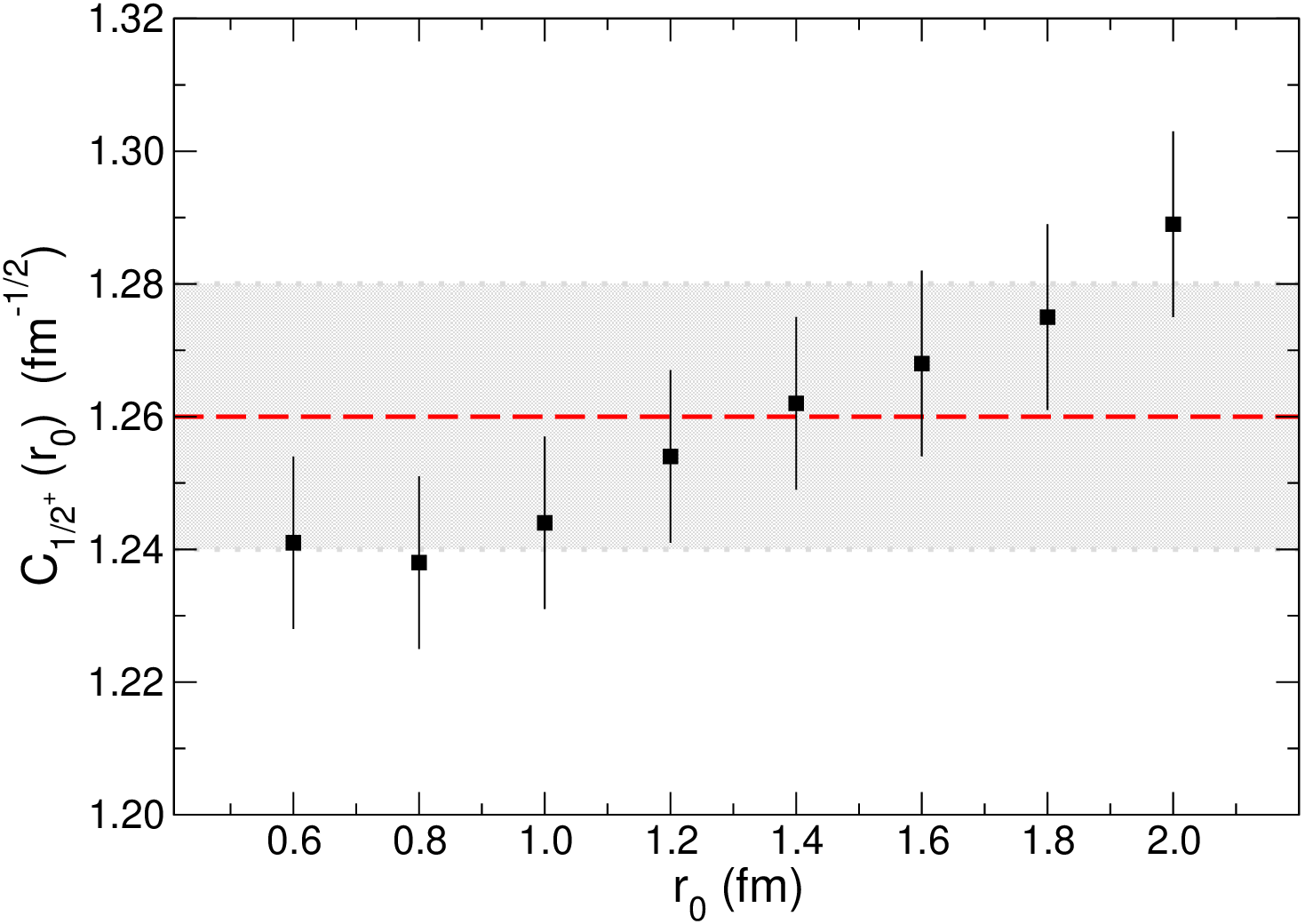}
\caption{\label{f-3} ANCs extracted for the $^{15}$C g.s.\ for each wave function of Fig.\ \ref{f-0_17MeV}.
Our recommended value is displayed by the horizontal red dashed line (the gray band represents its uncertainty).}
\end{figure}

Having determined the angular region within which the process is purely peripheral, we extract the value of the ANC ${\cal C}_{1/2^+}(r_0)$ for each of the single-particle wave function shown in \Fig{f-0_17MeV}.
This is done by scaling, through a  $\chi^2$ minimization, the corresponding theoretical cross section to the data selected at $\theta < 12^{\circ}$ \cite{yang-capel}. 
The ANCs ${\cal C}_{1/2^+}(r_0)$ obtained in this way are shown in \Fig{f-3} as a function of the potential width $r_0$.
The error bars correspond to the uncertainty in the $\chi^2$ minimization.
Despite the huge changes in the radial wave functions observed in \Fig{f-0_17MeV}, the ANCs extracted are nearly independent of $r_0$; they fall within 4\% from each other.
This is similar to what was obtained for $^{11}$Be (see Fig.~8 of \Ref{yang-capel}), hence confirming the validity of the method.

To deduce an estimate of the actual ANC ${\cal C}_{1/2^+}$, we average the ${\cal C}_{1/2^+}(r_0)$ results and get ${\cal C}_{1/2^+}=1.26\pm0.02$~fm$^{-1/2}$ (${\cal C}^2_{1/2^+}=1.59\pm0.06$~fm$^{-1}$) displayed as the horizontal red dashed line and gray band in \Fig{f-3}.
Following the same process, we obtain for the e.s.\ an estimate of the ANC of ${\cal C}_{5/2^+}=0.056\pm0.001$~fm$^{-1/2}$. 

We compare our estimate with values extracted from the analysis of other experiments in \tbl{tab3}.
Though on the lower end of the range, the ANC we obtain agrees with most of the others.
Our value is within the uncertainty band of the ANC extracted from knockout measurements in \Ref{15CANC_2002}, 
which is not surprising because that reaction is mostly peripheral \cite{HC19}.
Compared to the value extracted from the width of the $\half^+$ ground state of the proton-unbound mirror nucleus $^{15}$F, our ${\cal C}_{1/2^+}$ seems too low.
However, as explained in \Ref{Muk10}, that resonant state being quite broad, its width used in this analysis might be marred with significant uncertainty.
In \Ref{15CANC_2007}, Pang \etal\ have used the aforementioned $^{14}$C$(d,p)^{15}$C transfer data measured at $E_d=14$~MeV \cite{14MeVdp}, which have not enough points at forward angles to be purely peripheral.
Its large value is most likely due to that issue.
Note also that the normalization of the $E_d=14$~MeV data has been questioned in \Ref{17MeVdp}.
Interestingly, we are in excellent agreement with the value obtained by Summers and Nunes in their analysis \cite{SN08,SN08err} of the Coulomb breakup cross section of $^{15}$C measured at RIKEN \cite{Nak09-15C}.
Since this reaction is very peripheral \cite{CN07,CN17}, this is not surprising (see \Sec{breakup68AMeV}).
Our ANC is also perfectly compatible with the value extracted from the same data at $E_d=17.06$~MeV in \Ref{17MeVdp}.
The ${\cal C}_{1/2^+}$ we have obtained is on the lower end of the uncertainty range of the value extracted from the $^{13}{\rm C}(^{14}{\rm C},^{15}{\rm C})^{12}$C and $d(^{14}{\rm C},p)^{15}$C transfer experiments in \Ref{15CANC_2014}. % performed in inverse kinematics
However, these experiments have been performed at energies corresponding to $E_d\approx24$~MeV, where the reaction is not fully peripheral \cite{yang-capel}, which may explain the slight disagreement with our ANC.

\begin{table}[t]
\begin{tabular}{ccl}
 ${\cal C}_{1/2^+}^2$ (fm$^{-1}$)  & Ref. & Method\\
\hline
1.48 $\pm$ 0.18 & \cite{15CANC_2002} & Knockout\\ %agrees with us but on the lower side, probably if they do not include the excited state in their calculation that might affect the extraction of the ANC
1.89 $\pm$ 0.11  & \cite{15CANC_2006} & Mirror symmetry\\ %from /Gamma_p in 15F, but the experimental value depends on the way it is extracted, hence the difference? Mukhamedzhanov obtains a smaller width of the1/2+ resonant state in 15F, whih would lead to a smaller ANC.
 2.14  & \cite{15CANC_2007} & Transfer\\ %Use the data at 14MeV, which are not peripheral and hence might not be reliable
1.74 $\pm$ 0.11  & \cite{SN08,SN08err} & Coulomb breakup\\ %Slightly higher than us, but they consider a distorted p-wave, hence that reduces the breakup calculation and thus leads to an overestimation of the ANC (we have a good agreement with the bu data of RIKEN (and GSI) using the same ANC and assuming a phaseshift nil in the p waves
1.64 $\pm$ 0.26  & \cite{17MeVdp} & Transfer\\%In excellent agreement with our data, but we have much smaller uncertainty, probably because they extract the ANC from the entire angular range of the data (thy still say that it's the normalisation at fwd angle that matters
1.88 $\pm$ 0.18 & \cite{15CANC_2014} & Transfer\\%Higher than us, but not incompatible. Moreover, obtained from higher-energy reactions, which are not peripheral.
1.59 $\pm$ 0.06 & this work & Transfer\\
\hline
\end{tabular}
\caption{Comparison of ${\cal C}_{1/2^+}^2$ inferred for the $^{15}$C g.s.\ from various works.}
\label{tab3}
\end{table}

The value we have obtained from the method developed in \Ref{yang-capel} is therefore in good agreement with most of the values cited in the literature, and the differences we observe with previous analyses can be explained from uncertainties in these analyses.
Incidentally, as was observed in our previous analysis of the $^{10}{\rm Be}(d,p)^{11}{\rm Be}$ transfer \cite{yang-capel}, this ANC for the ground state of $^{15}$C is in excellent agreement with the ${\cal C}_{1/2^+}^2=1.644$~fm$^{-1}$ obtained by Navr\'atil \etal\ in the aforementioned \emph{ab initio} calculation of this one-neutron halo nucleus \cite{Nav18p}.
The present work will therefore provide a stringent test of the value predicted in that NCSMC calculation.

\subsection{Halo-EFT description of $^{15}$C at NLO\label{potentials}}
Having inferred a reliable value of the ANC for the $^{15}$C g.s., we can now proceed as suggested in \Ref{CPH18} and adjust a NLO Halo-EFT potential \eq{e-1_1} to describe this nucleus within our reaction models.
In the $s_{1/2}$ partial wave, the two depths of the Gaussian potential are fitted to reproduce the experimental binding energy of the halo neutron to the core and our ANC.
As in Refs.~\cite{CPH18,MC19}, we perform this fit for three different ranges $r_0$ to test the sensitivity of our reaction calculations to the short-range physics of the $^{14}$C-$n$ overlap wave function. The depths obtained by these fits are listed in Table \ref{tab4}.

\begin{table}[h!]
\begin{tabular}{c|cc|cc}
$r_0$ & $V_0^{s1/2}$ & $V_2^{s1/2}$ & $V_0^{d5/2}$ & $V_2^{d5/2}$ \\
(fm) & (MeV) & (MeV fm$^{-2}$) & (MeV) & (MeV fm$^{-2}$)\\
\hline
1.2 &-3.1995 & -71.3  & 169.299 & -92.368\\
1.5 &-92.814 & -2.70 &-91.000 & -9.000\\
2.0 &-80.827 & 2.70 &-94.916 & 2.53\\
\hline
\end{tabular}
\caption{Potentials describing $^{14}$C$+n$ g.s.\ and e.s.\ [see Eq.~(\ref{e-1_1})].
They are adjusted on the corresponding one-neutron binding energy and ANC.}
\label{tab4}
\end{table}

As mentioned earlier, the interaction in the $p$ wave is set to zero, in agreement with preliminary results of the \emph{ab initio} calculations \cite{Nav18p}.
In \tbl{tab4}, we also provide the depths for $^{14}$C-$n$ potentials in the $d_{5/2}$ partial wave, which are fitted to reproduce the binding energy and ANC of the $\fial^+$ excited bound state of $^{15}$C.
This goes beyond the NLO of Halo EFT, but it will enable us to check the influence of the presence of that state in the $^{15}$C spectrum in reaction calculations \cite{CPH18}.

\begin{figure}
\includegraphics[width=\columnwidth]{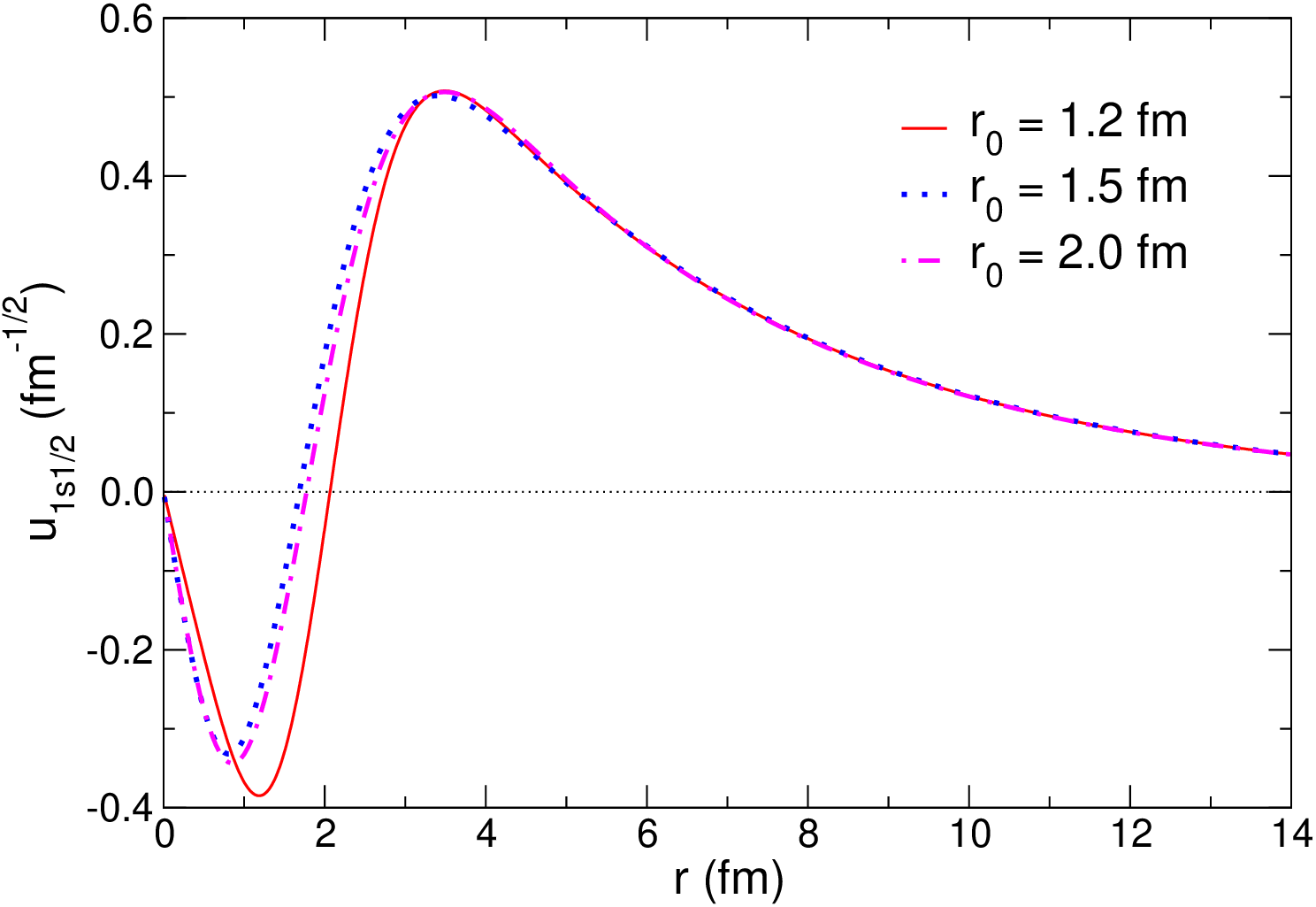}
\caption{\label{f-WF} 
Reduced radial wave functions of the  $^{15}$C g.s.\
obtained with the NLO Halo EFT potentials of \tbl{tab4}.}
\end{figure}

Figure \ref{f-WF} displays the $1s_{1/2}$ single-particle radial wave functions generated by the three potentials of \tbl{tab4}.
By construction, they exhibit the identical behavior in the asymptotic region, viz. for $r\gtrsim 4$~fm.
However, as expected, the three wave functions exhibit significant differences at short distances, which will enable us to test the sensitivity to the short-range physics of $^{15}$C of the various reactions we consider in the following.

\section{Transfer reaction $^{14}$C$(d,p)^{15}$C \label{transf-sec}}
We start our analysis of the reactions involving $^{15}$C using the NLO description developed in 
\Sec{potentials} by looking at how it behaves in transfer reactions.
We consider the low-energy reactions measured at $E_d=17.06$~MeV \cite{17MeVdp} and $E_d=14$~MeV \cite{14MeVdp}.
We use the same FR-ADWA model \cite{JT74} and potentials employed to extract the ANC in the previous section.

\begin{figure}
\includegraphics[width=\columnwidth]{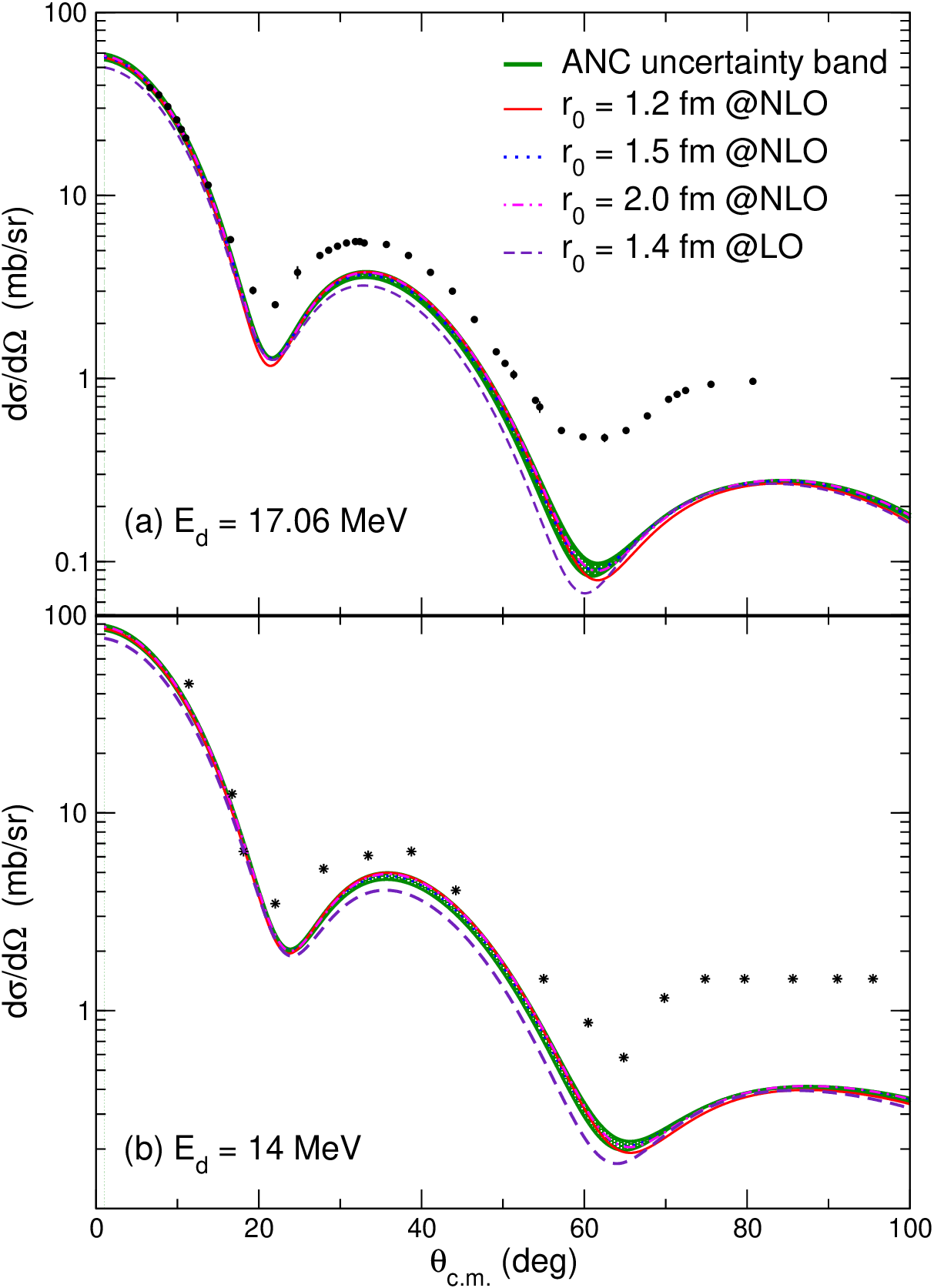}
\caption{\label{f-CS} 
Cross sections for the $^{14}{\rm C}(d,p)^{15}$C transfer reaction obtained at (a) $E_d=17.06$~MeV and (b) $E_d=14$~MeV.
FR-ADWA calculations performed with the NLO descriptions of $^{15}$C of \Sec{potentials} are compared to experimental data from (a) \Ref{17MeVdp} and (b) \Ref{14MeVdp}.
The green band shows the effect of the uncertainty on the ANC upon the calculation.}
\end{figure}

Figure~\ref{f-CS} displays the cross sections for the $^{14}{\rm C}(d,p)^{15}$C transfer reaction obtained at (a)~$E_d=17.06$~MeV and (b)~$E_d=14$~MeV.
The results of the FR-ADWA calculations for each of the three ranges of the Gaussian NLO potential \eq{e-1_1} are shown in the same colors and line types as the corresponding radial wave functions in \Fig{f-WF}.
The green band shows the uncertainty in the cross sections, obtained with the Gaussian potential of range $r_0=1.5$~fm, 
related to the uncertainty in the ANC we have extracted in \Sec{transfer}.
For comparison, we also show the results obtained with the LO description of $^{15}$C using $r_0=1.4$~fm (purple dashed line).

At $E_d=17.06$~MeV, without much surprise, the agreement of our NLO calculations with the data is perfect at forward angle since this is the region within which the fit has been performed in \Sec{transfer}.
The transfer cross section obtained with the LO description of $^{15}$C misses the data 
by a factor that corresponds to the value of the ANC, which is not fitted at this order. 
This confirms the importance of fitting both the energy and the ANC of the bound state to correctly reproduce the data.
All three NLO $^{14}$C-$n$ potentials provide the same cross section in the angular range of peripherality of the reaction, viz. $\theta<12^\circ$.
The agreement between the different wave functions actually extends beyond that range.
At larger angles, however, the transfer cross sections obtained with the three different single-particle $1s_{1/2}$ wave functions differ from one another, confirming that, at large angles, the reaction is sensitive to the short-range physics in $^{15}$C.
The uncertainty band encompasses the error bars of the forward-angle data, but cannot explain the discrepancy between our calculations and the experimental points at large angles.
This shows the limit of the present approach: Halo-EFT provides a proper low-energy---viz.\ large distances---description of the projectile, but, by construction, does not account for the details of the internal part of the $^{15}$C wave function.
Hopefully, including a more precise wave function of the projectile could improve the description of the data at large angles.
This could be done, e.g., using the overlap wave function provided by the \emph{ab initio} calculation of Navr\'atil \etal\ \cite{Nav18p}.
Alternatively, one could use a more elaborated two-body model of $^{15}$C, e.g., including core-excitation \cite{GMG15}.

\section{Coulomb breakup of $^{15}$C \label{sec3}}

We now turn to the Coulomb breakup of $^{15}$C.
As mentioned in the Introduction, this reaction has been measured on a lead target twice at two different energies.
First at GSI at 605~MeV/nucleon by Datta~Pramanik \etal\ \cite{Dat03} and second at RIKEN at 68~MeV/nucleon by Nakamura and his collaborators \cite{Nak09-15C}.
These two experiments are similar to those performed previously on the one-neutron halo nucleus $^{11}$Be \cite{gsi_exp,Fuk04}, which were recently successfully analyzed using a Halo-EFT description of  $^{11}$Be \cite{CPH18,MC19}.
We therefore follow these references and apply the same models of the reaction using the NLO description of $^{15}$C 
detailed in \Sec{potentials}.

\subsection{Breakup of $^{15}$C on lead at $605$~MeV/nucleon \label{breakup605AMeV}}

To analyze the breakup cross section of $^{15}$C measured on Pb at GSI at 605~MeV/nucleon \cite{Dat03}, we follow what we did in \Ref{MC19} and use an eikonal-based model of the reaction \cite{Glauber,BC12}, which properly accounts for special relativity.

In that model, the projectile $B$ is described by the two-body system introduced in \Sec{sec2}: a core $A$, to which a neutron $n$ is loosely bound, and which interact through the NLO Halo-EFT potential adjusted in \Sec{potentials}.
The target $T$ is seen as a structureless body of mass $m_T$ and charge $Z_Te$, which interacts with the projectile constituents $A$ and $n$ through the potentials $V_{AT}$ and $V_{nT}$, respectively.
We solve the problem within the Jacobi set of coordinates composed of the internal coordinate of the projectile $\ve{r}$ [see \Eq{e0}] and the relative coordinate of the projectile center of mass to the target $\ve{R}$.
The latter is explicitly decomposed into its longitudinal $Z$ and transverse $\ve{b}$ components relative to the incoming beam axis.

At this high beam energy, the use of the eikonal approximation is fully justified as well as the usual adiabatic---or sudden---treatment of the projectile dynamics during the reaction, i.e., we neglect the change in the projectile internal energy in comparison with its kinetic energy.
To properly account for special relativity, we follow Satchler \cite{satchler} and derive the eikonal wave function, which describes the projectile-target relative motion, from the Klein-Gordon equation expressed within the $B$-$T$ center-of-momentum (CM) frame \cite{satchler,pang}.
Within this description of the reaction, the three-body wave function exhibits the following asymptotic behavior
\beq
\Psi^{(m_0)}(\ve{R},\ve{r})\flim{Z}{+\infty} e^{iK_0Z} e^{i\chi(\ve{b},\ve{r})} \varphi_{n'_0l_0j_0m_0}(\ve{r}),
\eeqn{e1}
where $\hbar K_0$ is the initial $B$-$T$ momentum, $\chi$ is the eikonal phase that accounts for the interaction between the target and the projectile constituents, and $\varphi_{n'_0l_0j_0m_0}$ is the wave function of the projectile ground state, in which it is assumed to be initially.
Formally, the eikonal phase $\chi$ reads \cite{Glauber,BC12}
\beq
\chi(\ve{b},\ve{r})=-\frac{1}{\hbar v}\int_{-\infty}^\infty \left[ V_{AT}(\ve{R},\ve{r})+V_{nT}(\ve{R},\ve{r})\right]\,dZ,
\eeqn{echi}
where $v$ is the $B$-$T$ relative velocity.
This phase can be interpreted semi-classically by seeing the projectile $B$ following a straight-line trajectory at fixed impact parameter $\ve{b}$ along which its wave function accumulates a complex phase due to its interaction with the target.
It is composed of three terms: $\chi=\chi_{BT}^C+\chi^C+\chi^N$.
The first $\chi_{BT}^C(b)=2\eta\ln(K_0b)$, with $\eta=Z_BZ_Te^2/4\pi\epsilon_0\hbar v$, the Sommerfeld parameter of the reaction, simply describes the Coulomb scattering of the projectile by the target \cite{bertulani-danielewicz}.
It does not depend on $\ve{r}$, and hence does not contribute to the breakup of $B$.
The second
\beq
\chi^C(\ve{b},\ve{r})=\eta\int_{-\infty}^\infty \left(\frac{1}{\left|\ve{R}-\frac{m_n}{m_B}\ve{r}\right|}-\frac{1}{R}\right)\,dZ
\eeqn{echiC}
is the Coulomb term that contributes to the excitation of the projectile.
This phase diverges because the infinite range of the Coulomb interaction is not compatible with the sudden approximation, which assumes that the collision takes place in a short time.
To solve this issue, we use the Coulomb correction to the eikonal model (CCE) detailed in Refs.~\cite{bonaccorso-brink,CBS08}.
In that correction, the diverging eikonal Coulomb phase \eq{echiC} is replaced at the first order by the first order of the perturbation theory \cite{CBS08}
\beq
e^{i\chi^C}\rightarrow e^{i\chi^C}-i\chi^C+i\chi^{\rm FO}.
\eeqn{eCCE}
For the first-order estimate of the Coulomb phase, we consider the relativistic expression limited to the E1 term \cite{winther-alder}
\beq
\chi^{\rm FO}(\ve{b},\ve{r})=-\eta\frac{m_n}{m_B}\frac{2\omega}{\gamma v}\left[K_1\left(\frac{\omega b}{\gamma v}\right)\frac{\ve{b}\cdot\ve{r}}{b}+i\frac{1}{\gamma} K_0\left(\frac{\omega b}{\gamma v}\right)Z\right],
\eeqn{eFO}
where $\gamma=1/\sqrt{1-v^2/c^2}$ \footnote{Note the difference with \Ref{MC19}, where we had considered for the calculation of $\gamma$ the velocity of the projectile in the CM rest frame. 
Note also the correct formulation of our equation (\ref{eFO}) with the $1/\gamma$ factor (check Eq. (2.15) of Ref.~\cite{winther-alder}). 
These corrections have little effect on our results.}.

The third term of the eikonal phase $\chi^N$ corresponds to the nuclear interaction.
At low and intermediate energies, it is usually described by optical potentials fitted to reproduce elastic-scattering cross sections.
At high energy, and especially for exotic nuclei, it is difficult to find appropriate potentials.
Therefore, we rely on the optical limit approximation (OLA) of the Glauber theory \cite{Glauber,bertulani-danielewicz}, which has been successfully used in previous studies \cite{31Ne, MC19}.
In that approximation, the nuclear eikonal phase is obtained by averaging a profile function $\Gamma_{NN}$, which simulates the nucleon-nucleon interaction, over the density of the colliding nuclei
\beq
\chi^{OLA}_{xT}(\ve{b}_x)=  i \iint \rho_T(\ve{r'})  \rho_x(\ve{r''}) \Gamma_{NN}(\ve{b}-\ve{s'}+\ve{s''})d\ve{r''} d\ve{r'},
\eeqn{e8}
where $x$ stands for either $A$ or $n$, the two constituents of the projectile, and where $\ve{s'}$ and $\ve{s''}$ are the transverse components of the internal coordinate of the target ($\ve{r'}$) and $x$ ($\ve{r''}$), respectively.
In our three-body model of the reaction, the nuclear eikonal phase thus reads
\beq
\chi^N(\ve{b},\ve{r}) = \chi^{OLA}_{AT}(\ve{b}_A) + \chi^{OLA}_{nT}(\ve{b}_n).
\eeqn{e9}
We consider the usual form of the profile function
\beq
 \Gamma_{NN}(b)=\frac{1-i \alpha_{NN}}{4\pi \beta_{NN}}\sigma^{\rm tot}e^{-\frac{b^2}{2\beta_{NN}}}
\eeqn{e10}
where  $\sigma^{\rm tot}$ is the total cross section for the $NN$ collision,  $\alpha_{NN}$ corresponds to the ratio of the real to the imaginary part of the $NN$-scattering amplitude, and $\beta_{NN}$ is the slope of $NN$ elastic differential cross section.
These parameters are isospin dependent, which means that, in practice, the OLA phase \eq{e8} splits into four terms.
For the parameters of Eq.\ (\ref{e10}) we use the values provided in Ref.~\cite{OLA_prof-func_param} for an energy of 650~MeV.
The densities used in \Eq{e8} for the $^{14}$C core and the $^{208}$Pb target are approximated by the two-parameter Fermi distributions of Ref.~\cite{OLA_densities_param}, in which the authors study a systematization of nuclear densities based on charge distributions extracted from electron-scattering experiments as well as on theoretical densities derived from Dirac-Hartree-Bogoliubov calculations.
For $\rho_n$, we consider a Dirac delta function.

\begin{figure}
\includegraphics[width=\columnwidth]{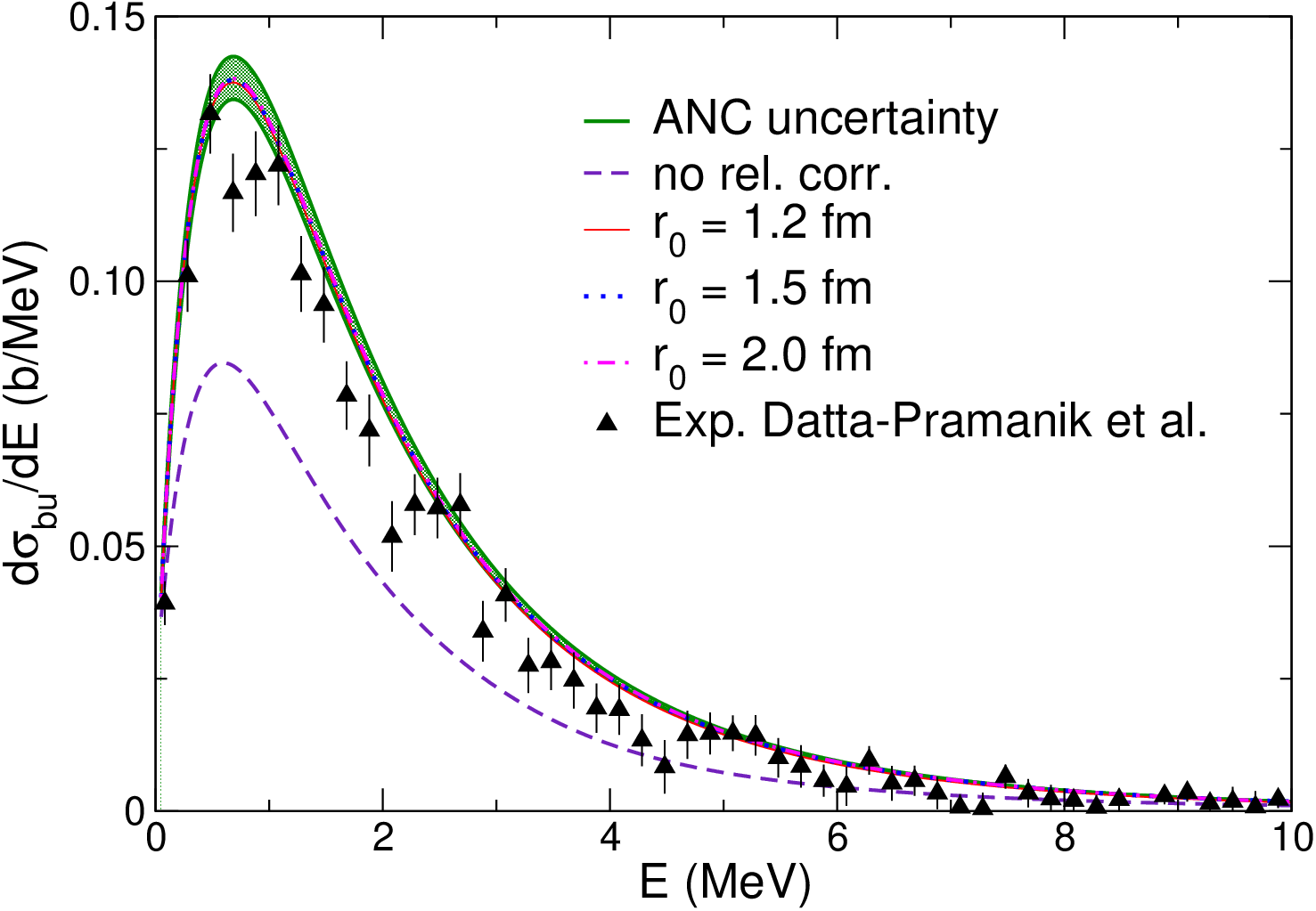}
\caption{\label{f2b} 
Breakup cross section of $^{15}$C on Pb at $605$~MeV/nucleon as a function of the relative energy $E$ between the $^{14}$C core and the neutron after dissociation. 
The results are obtained with the NLO Halo-EFT $^{14}$C-$n$ interactions listed in Tab.\ \ref{tab4}.
The green band represents the uncertainty on the $^{15}$C g.s.\ ANC.
For comparison with the GSI data of Ref.\ \cite{Dat03}, the theoretical predictions
have been folded with the experimental energy resolution \cite{gsi_exp}.
The result of the calculation without relativistic correction is shown as the purple dashed line.}
\end{figure}

The breakup cross sections obtained with this model of reaction are displayed in \Fig{f2b} as a function of the relative energy $E$ between the $^{14}$C core and the neutron after dissociation.
To enable the comparison with the experimental data of \Ref{Dat03}, all theoretical cross sections have been folded with the experimental energy resolution, which we have considered identical to the one provided by Palit \etal\ in the analysis of the Coulomb breakup of $^{11}$Be measured at GSI \cite{gsi_exp}.
The calculations performed with all three $^{14}$C-$n$ potentials listed in \tbl{tab4} are shown.
The sensitivity of our calculations to the uncertainty in the $^{15}$C g.s.\ ANC extracted in \Sec{transfer} is shown by the green band.
The result of the calculation obtained without relativistic corrections is displayed as the purple dashed line.
This clearly demonstrates the significance of these corrections at this beam energy.

Let us first note that our theoretical predictions are in excellent agreement with the data at all energies.
As expected, we do not note any appreciable difference between the calculations performed with the different Halo-EFT wave functions 
(see \Fig{f-WF}).
This result confirms that this reaction is purely peripheral, in the sense that it is sensitive only to the tail of the projectile wave function and not to its interior.
The excellent agreement with the data observed in this reaction observable suggests that the ANC we have extracted from the transfer data, combined with the choice of a nil interaction in the $p$ $^{14}$C-$n$ partial waves, is valid structurewise \cite{CN17}.
Accordingly, the predictions of the \emph{ab initio} calculations of Navr\'atil \etal\ seem correct \cite{Nav18p}.

In a subsequent test, we have analyzed how the inclusion of the $^{15}$C e.s.---described here as a $0d_{5/2}$ bound state (see \Sec{15Cstructure})---affects our breakup calculations.
The presence of that state in the $^{15}$C spectrum has no significant effect upon this reaction process; calculations performed with the Halo-EFT descriptions of $^{15}$C beyond NLO, which include this state, are nearly identical to those shown in \Fig{f2b}.
This is reminiscent of what has been observed in \Ref{CPH18} in the analysis of the RIKEN Coulomb breakup experiment of $^{11}$Be \cite{Fuk04}, in which the presence of the $\fial^+$ resonance, also described within the $d_{5/2}$ partial wave, is barely noticeable in the cross section.
This result is not surprising in a reaction that is strongly dominated by an E1 transition from the $s$ bound state towards the $p$ continuum.
The existence of a $d$ state in the low-energy spectrum of the projectile is more clearly seen in nuclear-dominated reactions, where quadrupole transitions are more significant \cite{CPH18,HC19}.
Therefore, for this Coulomb-dominated reaction, a Halo-EFT expansion limited to NLO is sufficient: 
the $d$ bound state would actually appear only at the next order (i.e.\ next to next to leading order, N$^2$LO), and it has nearly no influence in our breakup calculations.
This hence suggests that staying at NLO with a potential fitted to the ANC and binding energy of the g.s.\ in the $s$ wave and a nil potential in the $p$ wave, is enough to describe the experimental energy distributions for the breakup of $^{15}$C.

\subsection{Breakup of $^{15}$C on lead at $68$~MeV/nucleon \label{breakup68AMeV}}
The Coulomb breakup of $^{15}$C has also been measured on Pb at RIKEN at $68$~MeV/nucleon by Nakamura \etal\ \cite{Nak09-15C}.
To reanalyze these data using the Halo-EFT description of $^{15}$C developed in \Sec{potentials}, we consider the dynamical eikonal approximation (DEA) \cite{BCG05,GBC06}.
This model of reaction is also based on the eikonal approximation, however, it does not include the usual adiabatic approximation, which means that it properly includes the dynamics of the projectile during the collision, which has been shown to matter at this intermediate beam energy \cite{EBS05,CB05,SN08,Esb09}.
Besides having proved to be very efficient in the description of various observables measured in the breakup of one-neutron \cite{GBC06} and one-proton \cite{GCB07} halo nuclei, the model has been shown to be in excellent agreement with other breakup models on this very reaction \cite{CEN12}.

Following \Ref{CPH18}, we include the $^{14}$C-$n$ Halo-EFT potentials within the DEA and compute the breakup cross section at the RIKEN energy.
To describe the nuclear interaction between the projectile constituents and the target, we follow \Ref{CEN12} and consider optical potentials found in the literature.
The $^{14}$C-Pb potential is obtained from the scaling of an $^{16}$O-Pb potential fitted to reproduce the elastic-scattering cross section of these nuclei at 94~MeV/nucleon \cite{Rou88}.
We simply scale the radius of the potential by $0.987 = (14^{1/3} + 208^{1/3})/(16^{1/3} + 208^{1/3})$ to account for the mass difference between $^{16}$O and $^{14}$C and ignore the difference in beam energy.
We use the Bechetti and Greenlees global nucleon-target optical potential to simulate the $n$-Pb interaction \cite{BG69}.
Note that the details of these interactions are provided in the supplemental material of \Ref{CEN12}.

\begin{figure}
\includegraphics[width=\columnwidth]{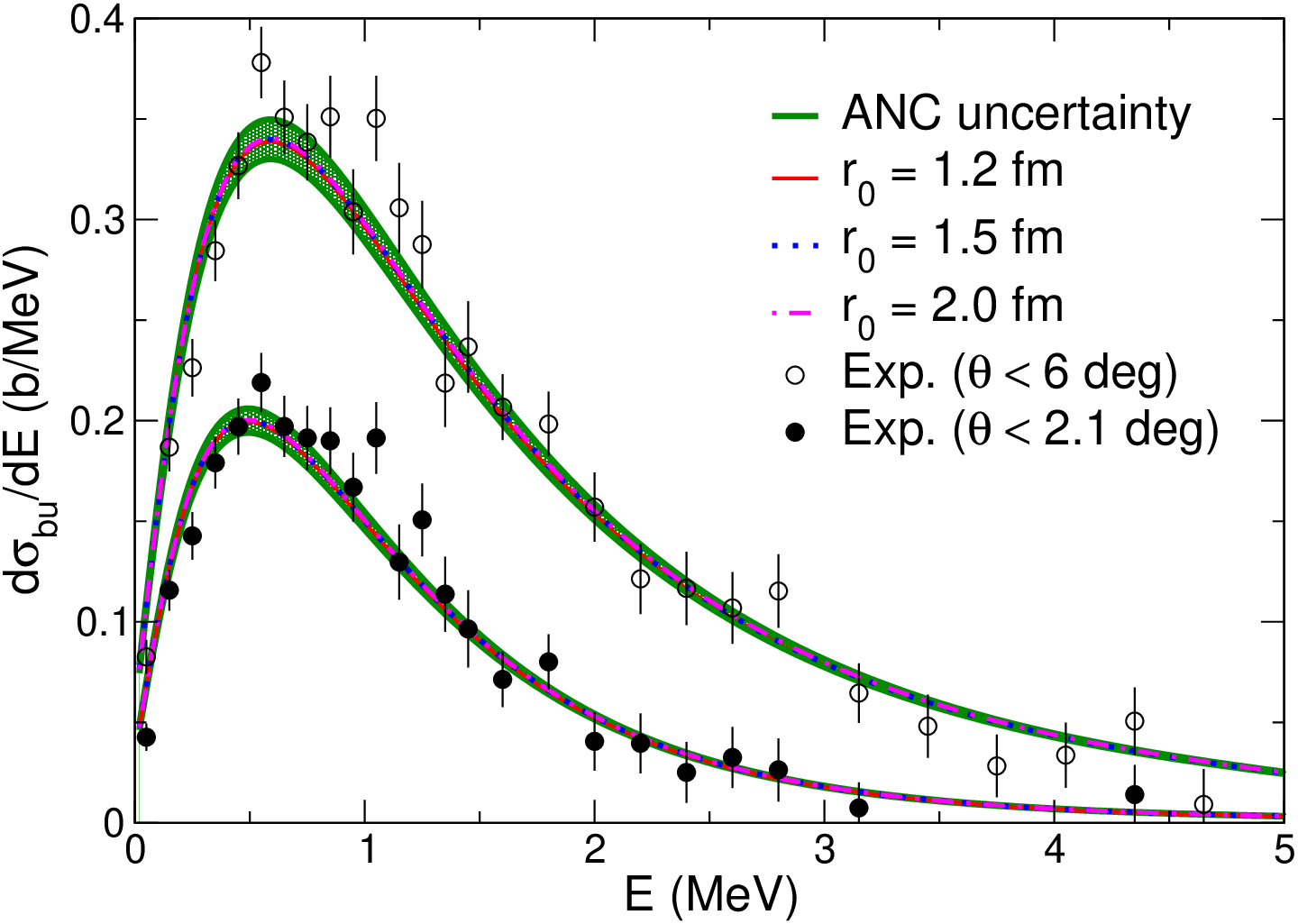}
\caption{\label{f4} Breakup cross section of $^{15}$C on Pb target at $68$~MeV/nucleon at two angular cuts plotted as a function of the relative energy $E$ between the $^{14}$C core and the neutron after dissociation. 
Results obtained with the different Halo-EFT $^{14}$C-$n$ interactions listed in \tbl{tab4} are shown. 
For comparison with the RIKEN data of Ref.~\cite{Nak09-15C}, the theoretical predictions have been folded with the experimental energy resolution.}
\end{figure}

The results of these calculations are shown in \Fig{f4} as a function of the $^{14}$C-$n$ continuum energy $E$.
We consider the two angular cuts under which the experimental data have been measured, i.e., $\theta<6^\circ$, which includes the entire significant angular range, and $\theta<2.1^\circ$, the forward-angle selection.
To allow for a direct comparison with the data of \Ref{Nak09-15C}, the results of our calculations have been folded with the experimental energy resolution.
The green band shows the effect of the uncertainty on the ANC. %we have extracted from the transfer data.

As in our analysis of the GSI experiment \cite{Dat03}, we obtain an excellent agreement with the data on the whole energy spectrum.
All three NLO $^{14}$C-$n$ potentials lead to identical cross sections showing that, at this energy also, the reaction is purely peripheral and that the ANC we have extracted from the low-energy transfer data and the nil phaseshift in the $^{14}$C-$n$ $p$ waves are consistent with this other set of data.
Our analysis hence independently confirms the value of the ANC extracted by Summers and Nunes from this same Coulomb breakup cross section \cite{SN08,SN08err}.
The slightly larger ANC they have obtained (see line 4 of \tbl{tab3}) is probably due to their use of a non-zero interaction in the $p$ wave, which tends to reduce these contributions to the breakup \cite{CN06,CN17,CPH18}.
Since there is no experimental observable upon which to constrain the phaseshift in these partial waves, we have to rely on theoretical hypotheses.
We have made a choice consistent with what we have done in the $^{11}$Be case \cite{CPH18} and with preliminary \emph{ab initio} predictions \cite{Nav18p}.
As shown in \Ref{CN17}, for the Coulomb breakup of loosely bound $s$ wave nuclei, it is the combination of ANC in the g.s.\ and phaseshift in the $p$ continuum that matters, especially at low energy $E$ in the $^{14}$C-$n$ continuum and forward scattering angle.
The excellent agreement with the data displayed in Figs.~\ref{f-CS}, \ref{f2b}, and \ref{f4} justifies our choice.
However, the uncertainty in the data is not sufficiently small to disprove the choice made in \Ref{SN08}.
Using their choice of $^{14}$C-$n$ potentials would most likely provide as good an agreement with experiment as ours.
Incidentally, this also confirms the \emph{ab initio} prediction of Navr\'atil \etal\ for the ANC of the $^{15}$C g.s.

In addition to these NLO calculations, we have also performed another set of calculations going beyond NLO by including the e.s.\ in the $^{15}$C spectrum as a $0d_{5/2}$ bound state.
The results, not shown here for clarity, are identical to those displayed in \Fig{f4}, confirming that in Coulomb-dominated reactions the details in the description of the $d$ waves are irrelevant, and that an NLO Halo-EFT description of the projectile is sufficient.

\section{Radiative capture $^{14}{\rm C}(n,\gamma)^{15}{\rm C}$\label{sec4}}

As mentioned in the Introduction, the radiative capture of a neutron by $^{14}$C to form a $^{15}$C nucleus [$^{14}{\rm C}(n,\gamma)^{15}{\rm C}$] plays a significant role in various astrophysical sites, from the possible inhomogeneous big-bang nucleosynthesis \cite{KMF90} to neutron-induced CNO cycles in AGB stars \cite{WGS99} and possible role in Type II supernov\ae\ \cite{Ter01}.
It is therefore useful for models of these astrophysical phenomena to have a reliable estimate of this reaction rate.
Unfortunately it is difficult to measure directly: both reactants are radioactive and, although $^{14}$C targets can be provided, obtaining purely monochromatic neutron beams is not simple.
This is why indirect techniques, such as the Coulomb-breakup method \cite{BBR86,BHT03}, have been proposed.
Nevertheless, recently, Reifarth \etal\ have taken up the gauntlet and performed a direct measure of this radiative capture \cite{Reifarth2008}.

In \Sec{sec3}, we have shown that the Halo-EFT description of $^{15}$C at NLO was sufficient to describe the breakup cross sections measured at GSI \cite{Dat03} and RIKEN \cite{Nak09-15C}.
As expected from the analyses published in Refs.~\cite{SN08,Esb09,CN17}, this model of $^{15}$C should also provide a good estimate for the radiative-capture cross section at low energy.
In this section, we compare our prediction with the data of Reifarth \etal\ \cite{Reifarth2008}.

The radiative-capture $^{14}{\rm C}(n,\gamma)^{15}{\rm C}$ is dominated by the E1 transition from the $p$ waves in the $^{14}$C-$n$ continuum towards the $1s_{1/2}$ ground state of $^{15}$C.
A small contribution comes also from the capture from the $p$ continuum waves to the $0d_{5/2}$ excited state of the nucleus.
Since these two contributions cannot be disentangled in the experiment of Reifarth \etal\ we use the Halo-EFT description of $^{15}$C beyond NLO to include this excited state in our model of the reaction.
To perform the calculations, we proceed as in \Ref{CN17}.

\begin{figure}
\includegraphics[width=\columnwidth]{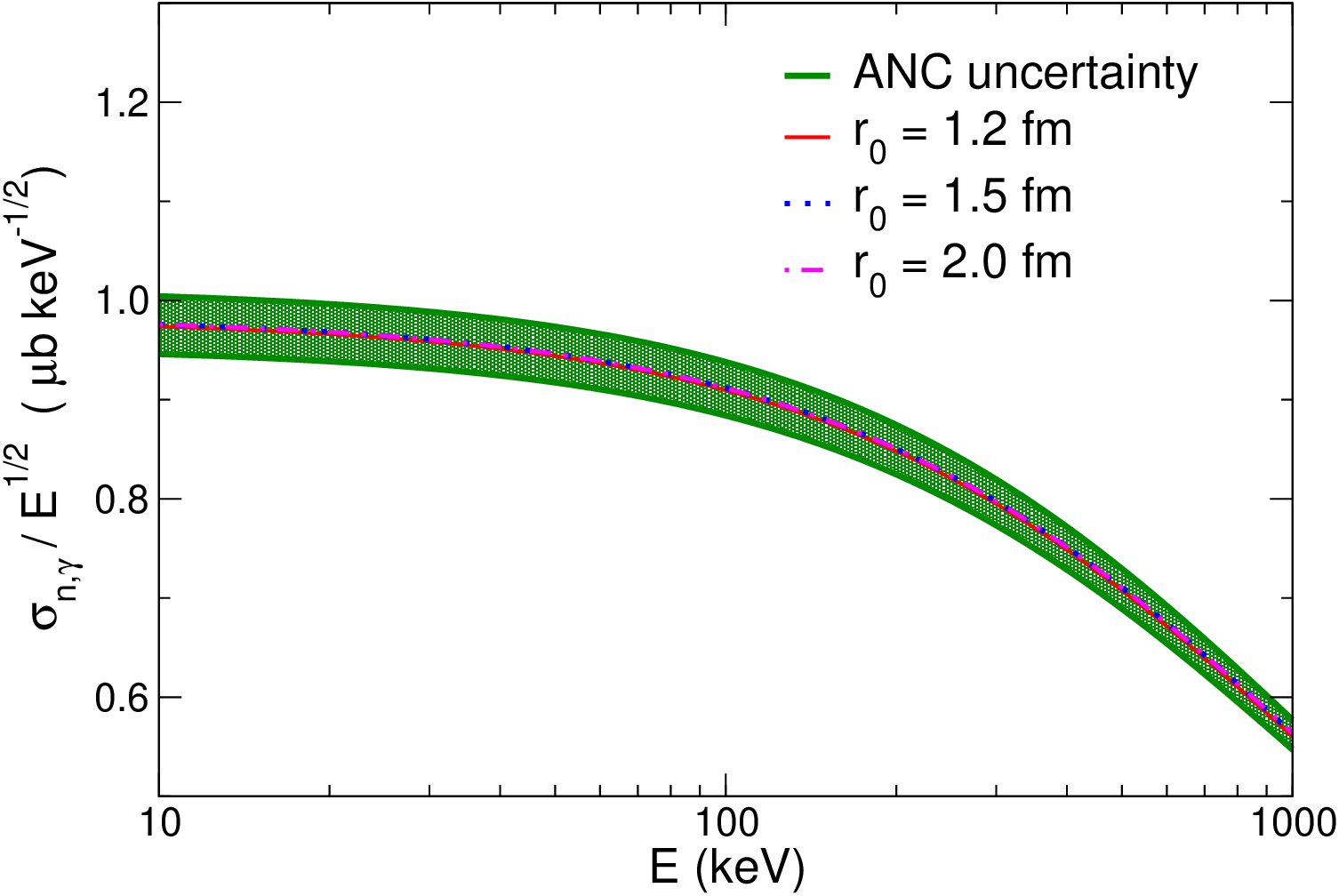}
\caption{\label{f5} Cross section for the radiative-capture $^{14}{\rm C}(n,\gamma)^{15}{\rm C}$.
The green band shows the uncertainty related to the ANC extracted from transfer data.}
\end{figure}

The radiative-capture cross section obtained in this way is displayed in \Fig{f5} as a function of the relative energy $E$ between the neutron and the $^{14}$C nucleus in the entrance channel.
The three $^{14}$C-$n$ Gaussian potentials provide identical cross sections, confirming that this reaction is purely peripheral \cite{15CANC_2006}.
The effect of the ANC uncertainty is shown by the green band.
The contribution due to the capture towards the $0d_{5/2}$ e.s.\ is, as observed elsewhere \cite{Reifarth2008,Esb09,CN17}, of the order of 5\%.
The details of the description of this state, and especially the accuracy of its ANC extracted from transfer data, are thus completely negligible in this analysis.
We have checked that the contribution of the E2 term to the radiative capture is orders of magnitude lower than the E1.
The cross section displayed in \Fig{f5} is in excellent agreement with prior predictions \cite{15CANC_2006,SN08err,RFV12,CN17} 
and the \emph{ab initio} prediction of Navra\'atil \etal\ \cite{Nav18p}.
It is however slightly lower than what has been obtained in the analysis of the direct experiment \cite{Reifarth2008}.

To properly confront these results with the data measured by Reifarth \etal\ \cite{Reifarth2008}, we need to account for the distribution of the neutron energy in the incoming beam \cite{CN18}.
The values averaged over the neutron distributions shown in Fig.~3 of \Ref{Reifarth2008} are provided in \tbl{tab5} alongside the experimental data.
\begin{table}
\begin{tabular}{c|cc}
$E$ (keV) & $\sigma_{n,\gamma}^{\rm exp}$ ($\mu$b) \cite{Reifarth2008} & $\sigma_{n,\gamma}^{\rm th}$ ($\mu$b)\\
\hline
23.3   & $7.1\pm0.5$   & $5.8\pm0.2$\\
150 & $10.7\pm1.2$ & $10.6\pm0.3$\\ 
500 & $17.0\pm1.5$ & $15.4\pm0.4$\\ 
800 & $15.8\pm1.6$ & $16.7\pm0.5$\\
\hline
\end{tabular}
\caption{Radiative-capture cross sections measured by Reifarth \etal\ \cite{Reifarth2008} and the theoretical results obtained with the Halo-EFT description of $^{15}$C developed in \Sec{potentials}. Our calculations include the small contribution of the capture to the excited $\fial^+$ of $^{15}$C described beyond NLO and are obtained after averaging over the energy distribution of the neutrons within the beams used in the experiment. The theoretical uncertainty corresponds to the uncertainty on the ANC we have extracted for the $^{15}$C g.s. The sensitivity to the choice of the range of the Gaussian potential $r_0$ is not seen at the level of precision displayed here.}
\label{tab5}
\end{table}
The experimental values are the ones provided in Table~V of \Ref{Reifarth2008}.
The theoretical cross sections are the one obtained using the $^{14}$C-$n$ potentials listed in \tbl{tab4} of the present article.
These values include the small contribution of the capture to the $0d_{5/2}$ bound state that simulates the $\fial^+$ e.s.\ of $^{15}$C.
The uncertainty provided for the theoretical value corresponds to the uncertainty on the ANC of the g.s.\ of $^{15}$C.
The sensitivity to the range $r_0$ of the Gaussian potential \eq{e-1_1} is smaller than the precision provided here.

Our theoretical predictions are usually in good agreement with the experimental values of Reifarth \etal\ \cite{Reifarth2008}.
The only significant difference is observed at the lowest energy point, where our prediction lies two sigma lower than the measured cross section.
This seems to be an issue for most of the indirect estimates of this cross section \cite{15CANC_2006,SN08err,Esb09,RFV12,CN17}.
Therefore, either there is some new physics not considered in the single-particle descriptions used in these references and in the present study, or there is some systematic uncertainty, which has not been well accounted for in the analysis of the experiment.
The cross section we derive from our Halo-EFT description of $^{15}$C at the single astrophysical energy $E=23.3$~keV is $\sigma_{n,\gamma}(23.3~{\rm keV})=4.66\pm0.14~\mu$b, which is slightly lower than what other groups obtain \cite{Reifarth2008,Tim14,SN08}.

\begin{figure}
\includegraphics[width=\columnwidth]{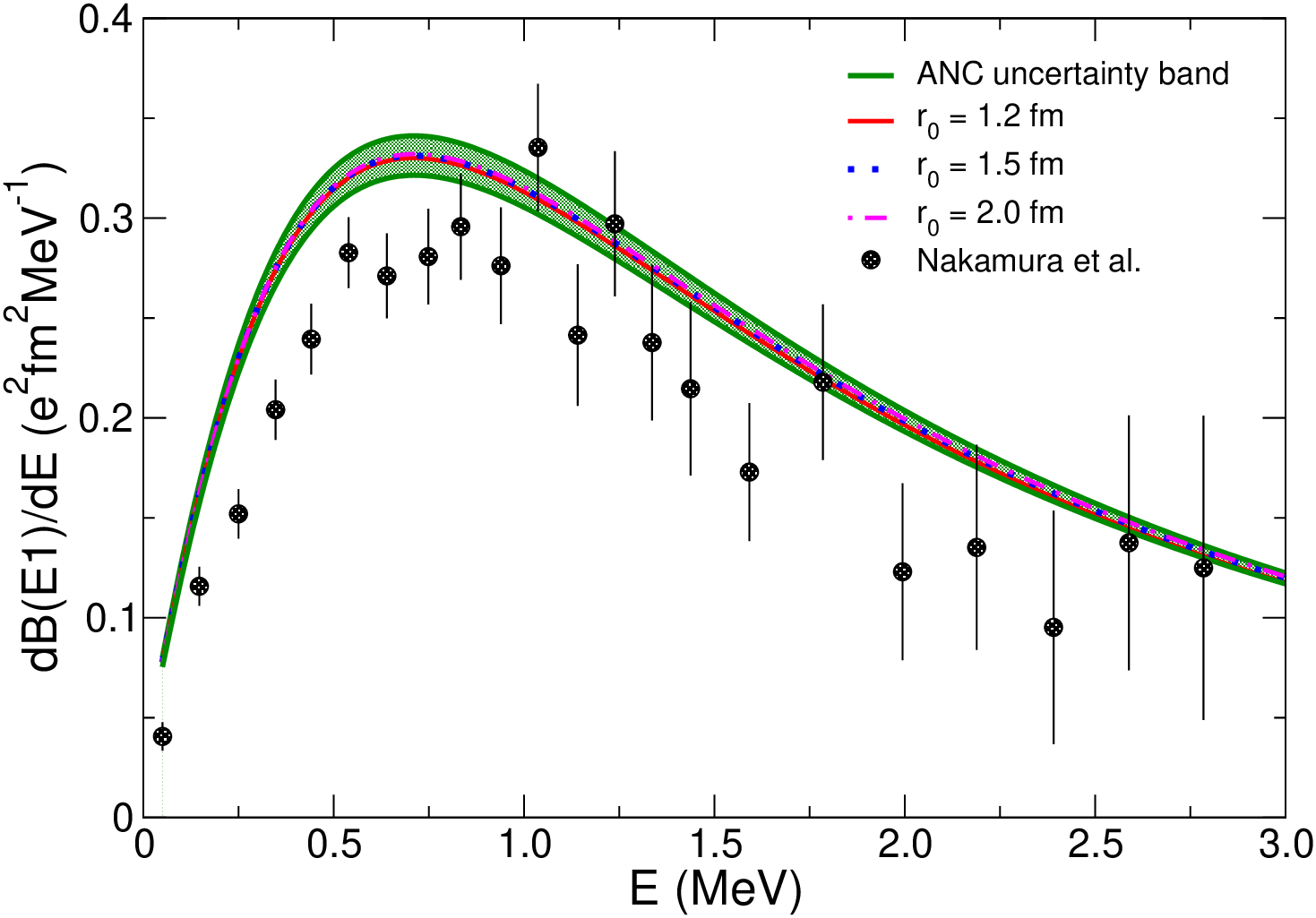}
\caption{\label{f-E1} Electric dipole strength deduced from the Halo-EFT structure of $^{15}$C at NLO, 
compared to the E1 strength inferred by Nakamura \etal~\cite{Nak09-15C}. 
For a better comparison, our calculation has been folded with the experimental resolution.}
\end{figure}

Within our study, this is the only one oddity in the analysis of various reaction observables, which are all peripheral, and in particular with Coulomb-breakup cross sections, which are sensitive to the same nuclear-structure observables as the radiative capture, viz.\ the ANC of the g.s.\ of $^{15}$C and the phaseshift in the $^{14}$C-$n$ $p$ waves \cite{CN17}.
We therefore believe that they are well constrained within our model of $^{15}$C.
The E1 strength this model predicts, and upon which both the Coulomb-breakup and the radiative-capture cross sections depend, should thus be quite reliable.
Figure~\ref{f-E1} provides this $dB({\rm E1})/dE$ as a function of the relative energy $E$ between the $^{14}$C
and the neutron in the continuum. The value we obtain from our NLO $^{14}$C-$n$ potentials are compared with the E1 strength 
inferred from the Coulomb-breakup measurement by Nakamura \etal~\cite{Nak09-15C}. 
We observe that the latter is systematically lower than the $dB({\rm E1})/dE$ deduced from our Halo-EFT model of $^{15}$C, 
even though we are in perfect agreement with their Coulomb-breakup cross sections (see Fig.~\ref{f4}). 
This difference is due to higher-order effects, which are neglected in the analysis of the RIKEN data. 
As already shown in Refs.~\cite{SN08,Esb09,CN17}, these effects are significant and cannot be ignored in the reaction model. 
This is the reason why the RIKEN prediction of the cross section for the radiative capture  $^{14}$C($n$,$\gamma$)$^{15}$C 
underestimates the direct measurement or Reifarth \etal~(see Fig.\ 3 of Ref.~\cite{Nak09-15C}). 
A comparison with that observable within the \textit{ab initio} model of  Navr\'atil \etal\ would be interesting to confirm our prediction.

\section{Summary and outlook \label{sec5}}
The exotic nucleus $^{15}$C raises interests in various fields.
It exhibits a one-neutron halo \cite{Tan96,Rii13}, and its synthesis through the radiative capture of a neutron by $^{14}$C takes place in various astrophysical sites \cite{WGS99,KMF90,Ter01}.
It is therefore interesting to better understand its structure and to provide astrophysicists with reliable cross sections for the radiative capture $^{14}{\rm C}(n,\gamma)^{15}{\rm C}$ at low energies.

In this work, we have reanalyzed various reactions involving $^{15}$C using one single description of that nucleus.
Following the work initiated in \Ref{CPH18}, we have considered a Halo-EFT description of that one-neutron halo nucleus.
Once coupled to a precise model of reactions, this very systematic expansion enables us to accurately determine the observables that affect the reaction process and hence, which can be probed through experimental measurements \cite{CPH18,yang-capel,MC19}.

Using a LO Halo-EFT Hamiltonian \eq{e-LO}, we have reanalyzed the $^{14}{\rm C}(d,p)^{15}{\rm C}$ transfer data at low energy \cite{17MeVdp} within the framework of the FR-ADWA \cite{JT74}.
Following the results of \Ref{yang-capel}, focusing on the forward-angle region, enables us to select purely peripheral data, from which a reliable estimate of the ANC of the g.s.\ of $^{15}$C has been inferred.
The value obtained ${\cal C}_{1/2^+}=1.26\pm0.02$~fm$^{-1/2}$ (${\cal C}^2_{1/2^+}=1.59\pm0.06$~fm$^{-1}$) is in good agreement with previous work \cite{15CANC_2002,15CANC_2006,15CANC_2007,SN08,SN08err,17MeVdp,15CANC_2014} and with preliminary \emph{ab initio} predictions \cite{Nav18p}.

The ANC hence obtained coupled to the binding energy of the valence neutron to the $^{14}$C provides us with two nuclear-structure observables, upon which we have constrained a Halo-EFT Hamiltonian at NLO.
This Hamiltonian has then be used within precise models of reactions to reanalyze transfer data \cite{14MeVdp,17MeVdp}, Coulomb-breakup cross sections measured at high \cite{Dat03} and intermediate \cite{Nak09-15C} energies, and cross sections for the radiative capture $^{14}{\rm C}(n,\gamma)^{15}{\rm C}$ \cite{Reifarth2008}.
In all cases, we observe a very good agreement with experiment without the need for any additional adjustment.

By showing that all these experiments can be described at the NLO of the Halo-EFT expansion, these analyses indicate that the core-neutron binding energy and the ground-state ANC are the sole nuclear-structure observables that need to be constrained to reproduce these data.
These reactions are therefore purely peripheral, 
in the sense that they probe only the tail of the projectile wave function and not its interior.
Especially, no need is found for a renormalisation of the projectile wave function, confirming that no spectroscopic factor can be extracted from such measurements \cite{CN07,CPH18}.
Going beyond NLO, we have found that the presence of the bound excited state of $^{15}$C in its description has no effect in Coulomb-breakup calculations.

From this NLO description of $^{15}$C we have been able to infer a reliable estimate of the E1 strength from the $\half^+$ ground state of $^{15}$C to its $^{14}$C-$n$ continuum.
This $dB({\rm E1})/dE$ leads to excellent agreement with the measurements of both the $^{15}$C Coulomb breakup \cite{Dat03,Nak09-15C} and the radiative capture $^{14}{\rm C}(n,\gamma)^{15}{\rm C}$ \cite{Reifarth2008}.
Accordingly, we suggest as a cross section for the latter process at astrophysical energy the value $\sigma_{n,\gamma}(23.3~{\rm keV})=4.66\pm0.14~\mu$b.

The excellent results obtained within this framework confirms the interest of coupling a Halo-EFT description of the nucleus to existing precise models of reactions \cite{CPH18}.
They also drive us to extend this idea to other reactions, like knockout \cite{HC19}.
Hopefully, the model developed herein and in \Ref{CPH18} will enable us to reproduce existing data on $^{15}$C and $^{11}$Be \cite{Tos02,Sau04,Fang04}.
We also plan to apply this model to other halo nuclei, like $^{19}$C and $^{31}$Ne.

\begin{acknowledgments}
This project has received funding from the European Union’s Horizon 2020 research and innovation programme under grant agreement No 654002,
the Deutsche Forschungsgemeinschaft within the Collaborative Research Centers 1044 and 1245, 
and the PRISMA (Precision Physics, Fundamental Interactions and Structure of Matter) Cluster of Excellence.
J.~Y.\ is supported by the China Scholarship Council (CSC).
P.~C.\ acknowledges the support of the State of Rhineland-Palatinate.
\end{acknowledgments}

%\bibliography{15C_bib.bib}
%apsrev4-2.bst 2019-01-14 (MD) hand-edited version of apsrev4-1.bst
%Control: key (0)
%Control: author (8) initials jnrlst
%Control: editor formatted (1) identically to author
%Control: production of article title (0) allowed
%Control: page (0) single
%Control: year (1) truncated
%Control: production of eprint (0) enabled
%
\end{document}